\documentclass[11pt]{article}

\usepackage{amsmath,amssymb,amsthm}
\usepackage{mathtools}
\usepackage{multirow}
\usepackage[a4paper]{geometry}
\usepackage{hyperref}
\usepackage{qcircuit}
\usepackage{xargs}
\usepackage{tikz}
\usepackage{qcircuit}
\usepackage{subcaption}

\usetikzlibrary{shapes.geometric, arrows}
\tikzstyle{startstop} = [rectangle, rounded corners, minimum width=3cm, minimum 
height=1cm,text centered, draw=black]
\tikzstyle{io} = [trapezium, trapezium left angle=70, trapezium right 
angle=110, minimum width=3cm, minimum height=1cm, text centered, draw=black]
\tikzstyle{process} = [rectangle, minimum width=3cm, minimum height=1cm, text 
centered, draw=black]
\tikzstyle{decision} = [diamond, minimum width=4cm, minimum height=1cm, text 
centered, draw=black]
\tikzstyle{arrow} = [thick,->,>=stealth]

\usepackage{authblk}
\usetikzlibrary{calc,positioning}
\usetikzlibrary{arrows,decorations.markings}

\newcommand{\ket}[1]{| #1 \rangle}
\newcommand{\bra}[1]{\langle #1 |}
\newcommand{\ketbra}[2]{| #1 \rangle \langle #2 |}
\newcommand{\braket}[2]{\bra{#1}{#2}\rangle}

\newcommand{\kron }{\otimes}

\newcommand{\Hl}{\mathcal{H}}
\newcommand{\Id}{\mathrm{I}}

\newcommand{\ie}{\emph{i.e.}}
\newcommand{\eg}{\emph{e.g.}}
\newcommand{\ii}{\mathrm i}

\newcommand{\suc}{\mathrm{suc}}
\newcommand{\R}{\mathbb R}

\newtheorem{definition}{Definition}
\newtheorem{theorem}{Theorem}
\newtheorem{corollary}{Corollary}
\theoremstyle{definition}
\newcommand{\cop}{\mathcal{C}}
\newcommand{\robber}{\mathcal{R}}
\newcommand{\doiLink}[1]{DOI: \href{http://dx.doi.org/#1}{#1}}
\newcommand{\qcircuitdots}{\push{\rule{-2.8em}{0em}\ldots \rule{.3em}{0em}}}

\title{The role of quantum correlations in Cop and Robber game}

\author{Adam Glos}
\author{Jaros{\l}aw Adam Miszczak}
\affil{Institute of Theoretical and Applied Informatics, Polish Academy of 
Sciences, Ba{\l}tycka 5, 44-100 Gliwice, Poland}

\date{05/11/2017 (v. 1.51)}

\begin{document}
\maketitle

\begin{abstract}
We introduce and study quantized versions of Cop and Robber game. We achieve
this by using graph-preserving quantum operations, which are the quantum
analogues of stochastic operations preserving the graph. We provide the tight
bound for the number of operations required to reach the given state. By
extending them to the controlled operations, we define a quantum-controlled Cop
and Robber game, which expands the classical Cop and Robber game, as well as the
classically controlled quantum Cop and Robber game. In contrast to the typical
scheme for introducing quantum games, we assume that both parties can utilise
full information about the opponent's strategy. We show that the utilisation of
the full knowledge about the opponent's state does not provide the advantage.
Moreover, the chances of catching the Robber decrease for classical cop-win
graphs. This result does not depend on the chosen model of evolution. On the
other hand, the possibility to execute controlled quantum operations allows
catching the Robber on almost all classical cop-win graphs. By this, we
demonstrate that it is necessary to enrich the structure of correlations between
the players' systems to provide a non-trivial quantized Cop and Robber game.
Thus the quantum controlled operations offer a significant advantage over the
classically controlled quantum operations.\\[4pt] \textbf{Keywords:}
combinatorial games, quantum networks, quantum entanglement.\\[4pt]
\textbf{MSC classes:} 05C57 (Primary), 91A46, 81P40 (Secondary).
\end{abstract}

%%%%%%%%%%%%%%%%%%%%%%%%%%%%%%%%%%%%%%%%%%%%%%%%%%%%%%%%%%%%%%%%%%%%%%%%%%%%%%%%
%\tableofcontents
%%%%%%%%%%%%%%%%%%%%%%%%%%%%%%%%%%%%%%%%%%%%%%%%%%%%%%%%%%%%%%%%%%%%%%%%%%%%%%%%

%%%%%%%%%%%%%%%%%%%%%%%%%%%%%%%%%%%%%%%%%%%%%%%%%%%%%%%%%%%%%%%%%%%%%%%%%%%%%%%%
\section{Introduction}
%%%%%%%%%%%%%%%%%%%%%%%%%%%%%%%%%%%%%%%%%%%%%%%%%%%%%%%%%%%%%%%%%%%%%%%%%%%%%%%%
Quantum information processing in complex networks is based on the assumption
that parties (or agents) acting in the network can utilize quantum carriers of
information to control the execution of protocols or algorithms. In this
scenario, it is reasonable to assume that the integrity of protocol execution
should be secure even against the attacker possessing the ability to operate on
quantum data. In other words, one has to revise the results concerning the
security of classical distributed protocols taking into account the quantum
model of computation~\cite{miszczak2012high}.
    
Suppose we have a reflexive graph $G$ (\ie\ with an edge joining a vertex to
itself at each vertex) and two players. We define a game as follows: the first
player, usually called \emph{Cop} or \emph{Pursuer}, chooses a vertex on which
he starts and then the second player, called \emph{Robber} or \emph{Evader},
chooses his vertex. After that, each player changes his position to a
neighbouring vertex, sequentially. If, after a finite number of rounds, the Cop
and the Robber meet in the same vertex, we say that the Cop wins the game.
Otherwise, we say that the Robber is the winner.

The game called \emph{Cop and Robber game} was first analysed by Quilliot
\cite{quilliot1978jeux} and Nowakowski and Winkler
\cite{nowakowski_vertex--vertex_1983}. The first author provided a beautiful
characterization of finite copwin graphs, \ie\ graphs on which the Cop has a
winning strategy. The other authors extended it to the case of infinite graphs.
Since then, many variants of the game have been proposed, including the game
where the players do not see each other~\cite{bernhard1987rabbit}, the players
move simultaneously \cite{konstantinidis2016simultaneously}, the Cop plays
heuristically \cite{bonato2016probabilistic,fitzpatrick2016deterministic}, or
the play is continuous in time \cite{fitzgerald1979princess,alpern2008princess}.
The game provides many stimulating mathematical problems. Among the best-known
of them are the characterization of the multi-copwin graph \cite{boyer2013cops}
and Meyniel conjecture~\cite{frankl1987cops}. Furthermore, the game provides an
interesting algorithmic problem: demonstrating that a graph needs at most $k$
Cops for winning the game is NP-hard. The game has also found its applications
in robotics \cite{chung2011search}, modelling graph searching
\cite{bonato2013graph}, and analysing the security in complex
networks~\cite{bonato2007pursuit}. For a survey of the game and its properties
see \cite{bonato2011game}.

In this work, we introduce the quantum games based on the assumption that both
parties possess full information about the strategy of the second party. In
quantum game theory, this assumption is rare as the main motivation to study
quantum games is to extend the space of possible moves
\cite{piotrowski2003invitation}, utilise entanglement for the synchronization of
the moves~\cite{eisert1999quantum}, or to share a quantum state for providing
the means for cooperation~\cite{pawela2013cooperative}. Such an approach was
used in \cite{hao2013novel} to analyse the quantum Prisoner's Dilemma as a
pursuit game on a graph, where the Cops utilise a shared entangled state.

In contrast, this paper contributes to the field of combinatorial games, where
all players (or parties) have full knowledge of the state of the system. In the
case of quantum combinatorial games, this is to say that both players possess
the complete knowledge of the previous moves of the opponent. In other words, we
are interested in the existence of the strategies, which are optimal in some
sense. The properties of such games, defined in quantum information systems,
were previously studied in the context of network exploration
problems~\cite{miszczak2014quantum} and modelling of trapping mechanism in
quantum networks \cite{sadowski2016lively}. Moreover, the possibility of using
the superposition of moves in a combinatorial game has been
considered~\cite{dorbec2017quantum}.

Here we consider the possibility to define a quantum extension of Cop and Robber
game, which describes the general protocol for tracking (or controlling) mobile
agents in the scenario with dynamical information. The main question arising in
this context concerns the possibility of constructing a strategy for the Cop to
catch the Robber, taking into account the probabilistic nature of the quantum
measurement. We analyse this problem by introducing suitable winning conditions.

The main contribution of the presented paper is obtaining a nontrivial quantum
version of the game. We propose and investigate graph-preserving quantum
evolution, which can be used in extending other graph-based games \eg\ simple
stochastic games~\cite{condon1990algorithms}. We prove the tight bound for the
number of required operations. Moreover, we show that quantum controlled gates,
and hence quantum entanglement, provide new strategies which diametrically
change the course of the game. Consequently, we provide an argument confirming
the crucial role of entanglement in quantum information processing.

The paper is organised as follows. In Section~\ref{sec:preliminaries}, we
introduce mathematical apparatus, including the concepts necessary to deal with
the quantum measurement. We achieve this by introducing a classical game called
open probabilistic Cop and Robber game. We show that the game trivialises in the
sense of the winning strategy. In Section~\ref{sec:evolution}, we define a
graph-preserving quantum evolution, and by applying it, we introduce classically
controlled quantum Cop and Robber game. We show that such a game cannot be used
to obtain an advantage in the sense of the broader class of winning strategies.
We also provide a tight bound for state obtainability. In
Section~\ref{sec:the-game}, we extend the graph-preserving quantum evolution to
describe the non-trivial extension of the game, namely quantum controlled Cop
and Robber game. We demonstrate that such version provides an advantage over the
classically controlled quantum scheme. Finally, in
Section~\ref{sec:conclusions}, we provide a summary of the presented results.

%%%%%%%%%%%%%%%%%%%%%%%%%%%%%%%%%%%%%%%%%%%%%%%%%%%%%%%%%%%%%%%%%%%%%%%%%%%%%%%%
\section{Preliminaries} \label{sec:preliminaries}
%%%%%%%%%%%%%%%%%%%%%%%%%%%%%%%%%%%%%%%%%%%%%%%%%%%%%%%%%%%%%%%%%%%%%%%%%%%%%%%%

We start by introducing mathematical apparatus required for the analysis of the
quantum Cop and Robber game. In particular, we provide some facts related to the
probabilistic version of the game that offer some insights into the behaviour of
the players in the quantum realm.

%%%%%%%%%%%%%%%%%%%%%%%%%%%%%%%%%%%%%%%%%%%%%%%%%%%%%%%%%%%%%%%%%%%%%%%%%%%%%%%%
\subsection{Graph terminology}
%%%%%%%%%%%%%%%%%%%%%%%%%%%%%%%%%%%%%%%%%%%%%%%%%%%%%%%%%%%%%%%%%%%%%%%%%%%%%%%%
Suppose we have a digraph $G$ with vertex set $V$ and arcs $A\subset V\times V$.
We say that the vertex $v\in V$ is a neighbour of $u\in V$ iff $(u,v)\in A$. The
set of all neighbours of $u$ is denoted by $S(u)$. A graph is called reflexive,
iff for each $v\in V$ we have $(v,v)\in A$. The reflexive graphs are the only
ones considered here. We say that the vertex $v$ is a corner if there exists the
vertex $u$ such that $S(v)\subset S(u)$. A spanning tree is a sub-graph that is
a tree which includes all of the vertices of $G$. A directed graph is called
reversible, if for arbitrary $v,w\in V$ there is path from $v$ to $w$.

Suppose we have an undirected graph $G$ with vertex set $V$ and edges $E$. We
call $V_1\subset V$ a dominating set iff for an arbitrary vertex $v\in V$ we
have $v\in V_1$ or $N(v)\cap V_1\neq \emptyset$, where $N(v)$ denotes the set of
neighbours of $v$. The vertex which itself forms a dominating set is called
\emph{universal vertex}.

A graph homomorphism from a graph $G$ to $G'$ is an arc-preserving mapping from
$V$ to $V'$. We call a graph homomorphism $f$ a retraction if for each $v'\in
V'$ we have $f(v')=v'$. Then we call $G'$ a retract of $G$ and denote it by
$G\upharpoonright V'$.

%%%%%%%%%%%%%%%%%%%%%%%%%%%%%%%%%%%%%%%%%%%%%%%%%%%%%%%%%%%%%%%%%%%%%%%%%%%%%%%%
\subsection{Open probabilistic Cop and Robber game}
%%%%%%%%%%%%%%%%%%%%%%%%%%%%%%%%%%%%%%%%%%%%%%%%%%%%%%%%%%%%%%%%%%%%%%%%%%%%%%%%
As it has already been mentioned, many variations of the game can be proposed.
One of the most popular versions is Hunter and Rabbit game
\cite{bernhard1987rabbit}, where players do not see each other until they are at
the same vertex. In that case, the strategy needs to be described with the
stochastic operations, and the players' positions are represented with
probability vectors. Note that the probabilistic version of Cop and Robber game
is no longer open since both the positions of the players, even in the sense of
probability vector, and the performed stochastic operations are unknown to the
opponent.

To introduce the open probabilistic version of Cop and Robber game, we use the
following scheme. Suppose that the graph is known to both players. The Cop and
the Robber, in that order, choose their initial positions randomly, \ie{}~they
select the probabilistic vector of a position. Next, they sequentially perform
stochastic operations preserving the graph structure on their states. In each
step, both the Cop and the Robber do not know their position and their
opponent's position, but they know the current probability of a player being in
all vertices. The Cop when to perform a measurement, that is when he uncovers
the board.

To consider probabilistic versions of Cop and Robber game, including the variant
with quantum strategies, we need to take into account the probabilistic nature
of quantum measurement. To this end, we define the following classes of winning
conditions.

\begin{definition}[$p$-copwin graphs]
We say that the graph is {$p$-copwin}, if there
is such strategy for the Cop that after a finite number of steps the probability of measuring the
players in the same vertex is greater than $p$. 
\end{definition}

\begin{definition}[Nearly $p$-copwin graphs]
We say that the graph is {nearly
$p$-copwin}, if for arbitrary $\varepsilon>0$ there is a strategy for the Cop such
that he can win with the probability at least $p-\varepsilon$. 
\end{definition}

It is worth noting that in the sense of the strategy set, the probabilistic
version expands the original, deterministic Cop and Robber game. Every move in
the original one can be described as a stochastic operation. When we restrict
players to deterministic probability vectors, each course of the original game
can be described with open probabilistic Cop and Robber formalism.

However, it is easy to see that the open probabilistic model narrows the set of
feasible strategies. To demonstrate this let us suppose that directed graph
$G=(V,A)$ is given. The Cop chooses vector $p_{\cop}(v) =\frac{1}{|V|}$ as the
initial state. For arbitrary Robber's position $p_\robber$ the probability of
measuring the players in the same vertex equals
\begin{equation}
p_{\textrm{copwin}} = \sum_{v\in V}p_\cop (v)p_\robber (v) = 
\frac{1}{|V|}\sum_{v\in V} p_\robber(v) = \frac{1}{|V|}.
\end{equation}

Note that the probability of winning with this strategy does not depend on the
Robber's position. Hence, an arbitrary directed graph is at least
$\frac{1}{|V|}$-copwin. If the Robber chooses the same initial state, we can
show that the arbitrary graph is precisely $\frac{1}{|V|}$-copwin. Note that the
strategy does not depend on the arcs set, and henceforth the graph does not need
to be even connected. The probability depends on the size of $V$ only.

%%%%%%%%%%%%%%%%%%%%%%%%%%%%%%%%%%%%%%%%%%%%%%%%%%%%%%%%%%%%%%%%%%%%%%%%%%%%%%%%
\subsection{Probabilistic versus deterministic strategy}
%%%%%%%%%%%%%%%%%%%%%%%%%%%%%%%%%%%%%%%%%%%%%%%%%%%%%%%%%%%%%%%%%%%%%%%%%%%%%%%%

Let us now analyse how the introduced scheme influences the class of cop-win
graphs. To achieve this let us examine an unfair game on cop-win graphs. It is
easy to see that if the Robber is limited to deterministic strategies while the
Cop is allowed to use randomised strategies, the latter can win with probability
arbitrarily close to one. The following theorem gives the full characterisation
of this situation.

\begin{theorem}
In open probabilistic Cop and Robber game on an undirected graph $G$ with the
Robber using deterministic moves, the Cop can win with probability one if and
only if $G$ is cop-win. Otherwise, the Cop can win with the probability
arbitrarily close to one.
\end{theorem}

\begin{proof}

The sufficiency of a graph being copwin comes from the fact that the Cop can use
the deterministic strategy to win. In the not cop-win graph case, the Robber can
always evade a part of the probability of the Cop localised in some vertex.
Hence we have necessity.

Now we show that an arbitrary graph is nearly 1-copwin. The Cop spreads uniquely
on the arbitrary dominating set of the graph. Next, he can `catch' the evader
with probability one over the cardinality of the dominating set. The part of
probability which is on the Robber's position will now follow him, while the
rest of the probability repeat the strategy. Each round needs the linear time of
single steps on a number of vertices, and one can show that the probability of
winning for the Robber decreases geometrically with the number of rounds. 
\end{proof}

If we swap the limitations, the Cop can win with probability at most
$\frac{1}{|V|}$, since the Robber can choose the uniform distribution as the
initial state. However, if a graph is not copwin, the Robber can use
deterministic strategies to avoid the pursuer entirely. We have not found the
full characterization of this situation for copwin graphs.

%%%%%%%%%%%%%%%%%%%%%%%%%%%%%%%%%%%%%%%%%%%%%%%%%%%%%%%%%%%%%%%%%%%%%%%%%%%%%%%%
\section{Graph-preserving discrete evolution} \label{sec:evolution}
%%%%%%%%%%%%%%%%%%%%%%%%%%%%%%%%%%%%%%%%%%%%%%%%%%%%%%%%%%%%%%%%%%%%%%%%%%%%%%%%

The first quantum version of the Cop and Robber game considered in this paper is
defined by using quantum operation, preserving the structure of a graph. In this
section, we use such operations to introduce a direct quantum version of the
open probabilistic Cop and Robber game. We call this game the \emph{classically
controlled quantum Cop and Robber game} to emphasise the fact that the
introduced scheme is based on the full knowledge of the quantum state of the
opponent. We show that in the context of winning strategy the game does not
change in comparison to the probabilistic version. We also show that for
arbitrary connected, reflexive and undirected graph every quantum state is
obtainable in $O(|V|)$ steps, which is tight bound. We also provide examples
demonstrating that finding a graph preserving discrete evolution is not always
trivial and can lead to some unintuitive results.

%%%%%%%%%%%%%%%%%%%%%%%%%%%%%%%%%%%%%%%%%%%%%%%%%%%%%%%%%%%%%%%%%%%%%%%%%%%%%%%%
\subsection{Direct approach}
%%%%%%%%%%%%%%%%%%%%%%%%%%%%%%%%%%%%%%%%%%%%%%%%%%%%%%%%%%%%%%%%%%%%%%%%%%%%%%%%
In the probabilistic version of Cop and Robber game, the players can choose
arbitrary stochastic operations preserving the graph structure. Such operations
are defined as follows.

\begin{definition}[Graph preserving stochastic operation]
We say that the stochastic operation $M$ preserves the graph, if for arbitrary
disconnected vertices $v$ and $w$, represented by orthogonal states $\ket{v}$
and $\ket{w}$ respectively, we have $\bra w M \ket v=0$.
\end{definition}

In a similar manner, one can define a quantum operation preserving the graph
structure.

\begin{definition}[Graph preserving quantum operation]
We say that the unitary operator $U$ is graph preserving quantum operation, if
for arbitrary disconnected vertices $v$ and $w$, represented by orthogonal
states $\ket{v}$ and $\ket{w}$ respectively, we have $ \bra w U \ket v =0$.
\end{definition}

Using the above definition one can introduce the quantized version of Cop and
Robber game, called \emph{classically controlled quantum Cop and Robber game}.
At the beginning of this game, both players choose ar arbitrary pure quantum
state. The evolution of pure states is described with graph preserving quantum
operation $U$. The $\mathbf U_G$ denotes the set of all quantum operations
preserving the graph $G$. The $\mathbf U_G$ contains the identity and is closed
under Hermitian transposition. However, it does not form a group because it is
not closed under multiplication.

\begin{figure}[t!]
	\begin{center}
		\mbox{ \Qcircuit @C=-0.5em @R=-0.5em @! {
				& \push{\rule{-2.8em}{0em}\qquad\quad
					\ket{\robber} \rule{.3em}{0em}} & \control\qw\cwx[1] & \gate {U_{\robber,1}} &
				\control  \qw\cwx[1] & \gate {U_{\robber,2}} & \qw & \qcircuitdots & \gate
				{U_{\robber,t-1}} &  \control\qw\cwx[1] & \meter &\cw \\
				\lstick{\ket{\cop}} & \qw\cwx[-1] & \gate{U_{\cop,1}} &\control   \qw\cwx[-1] &
				\gate {U_{\cop,2}} & \control   \qw\cwx[-1]  & \qw &\qcircuitdots & \control  
				\qw\cwx[-1]  & \gate {U_{\cop,t}} & \meter &\cw \\ }}
	\end{center}
	\caption{Classically controlled quantum Cop and Robber game. The Cop chooses
	his initial state first, and then the Robber chooses his state. Double line
	denotes classical control since both parties possess the full knowledge of
	their and their opponent's current state. The measurement is performed in the
	basis $\{\ket{v} : v \in V \}$.}
	\label{fig:classically-controlled-game}
\end{figure}
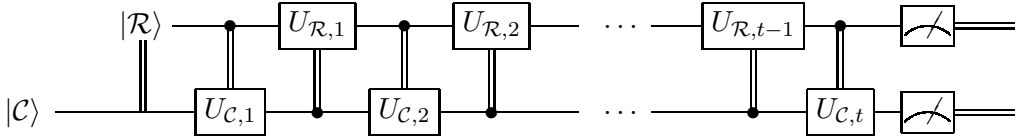

More precisely, we can define the game as follows. Each player has their own
system $\Hl_\robber $ and $\Hl_\cop$ spanned by orthogonal set $\{\ket v :v\in
V\}$ and chooses the initial quantum state $\ket {\robber_0}$ and $\ket
{\cop_0}$. The combined system is of the form $\Hl_\robber \kron \Hl_\cop$ and
the initial state is of the form $\ket{\robber_0}\kron \ket {\cop_0}$. Then, in
each iteration the Cop and the Robber perform operations $\Id_\robber\kron
U_\cop$ and $U_\robber\kron \Id_{\cop}$, respectively, where
$U_\cop,U_\robber\in\mathbf U_G$. In the end, they perform the measurement in
basis $\{\ket v: v\in V\}$. The Cop wins if, after $t$ steps and his move, both
players are measured in the same vertex. In this situation, if $\ket{S_t}$
denotes the game state after $t$ steps, then the probability of the Cop being a
winner reads
\begin{equation}
p_{\mathrm{copwin}} = \sum_{v\in V} |(\bra v \kron \bra v)\ket {S_t}|^2.
\end{equation}
One should note that by local operation one cannot entangle the registers. For
this reason the state is of the form $\ket{S_t} = \ket{\robber_t}\kron
\ket{\cop_t}$ and the formula above simplifies to
\begin{equation}
p_{\mathrm{copwin}} = \sum_{v\in V} |\braket{v}{\robber_t}\braket{v}{\cop_t}|^2.
\end{equation}
Quantum circuit for the above game is presented in Fig.~\ref{fig:classically-controlled-game}.

The quantized version of Cop and Robber game introduced above can be understood
as a reformulation of the open probabilistic Cop and Robber game in the language
of quantum computing. It can be seen that the ability to utilise local quantum
moves does not provide any party with an advantage. To demonstrate this one can
observe that the Cop and the Robber can choose the equal superposition of the
base states, $\frac{1}{|V|} \sum_{v\in V} \ket{v}$, as the initial state. It is
easy to observe that every graph is $\frac{1}{|V|}$-copwin. One should note that
this example suggests that the triviality does not depend even on the evolution
form -- similar results will be obtained if we choose discrete quantum walk
model or other models as an evolution.

The unfair game yields the same results as an open probabilistic game. We can
conclude the results as follows: \emph{in order to provide interesting,
non-trivial quantized Cop and Robber game, we need to enrich the structure of
correlations between the players' systems}. This motivates the quantum model
based on the ability to execute non-local quantum gates, developed in
Section~\ref{sec:the-game}.

%%%%%%%%%%%%%%%%%%%%%%%%%%%%%%%%%%%%%%%%%%%%%%%%%%%%%%%%%%%%%%%%%%%%%%%%%%%%%%%%
\subsection{Tight bound for state obtainability}
%%%%%%%%%%%%%%%%%%%%%%%%%%%%%%%%%%%%%%%%%%%%%%%%%%%%%%%%%%%%%%%%%%%%%%%%%%%%%%%%

It was shown \cite{montanaro2007quantum} that in order to evolve using
operations from $\mathbf U_G$ the graph may be directed but it must be
reversible, \ie{}~if there is a path from $v$ to $w$, then there is a path from
$w$ to $v$. However, it is not enough for obtaining an arbitrary result state.
Suppose $G=(\{0,\dots,n-1\},A)$ is a directed, clockwise cycle. Then, $\mathbf
U_G$ consist of identity operations and clockwise permutations changing at most
the amplitude phase, \ie{} 
\begin{equation}
U = \sum_{i=0}^{n-1} e^{\ii\alpha_i}\ketbra{i\oplus 1}{i},
\end{equation}
where $\oplus$ denotes addition modulo $n$, and $\alpha_i\in\R$. As an example,
if a state $\ket 0$ is given, it is impossible to obtain superposition
$\frac{1}{\sqrt{2}}(\ket{0}+\ket 1)$. Below we present a sufficient condition
for an arbitrary state obtainability.

\begin{theorem}\label{th:tight}
Let $G=(V,E)$ be such a reflexive, reversible digraph that contains an 
undirected spanning tree. 
Then, for 
arbitrary states $\ket{\varphi}$ and $\ket{\psi}$, there exists a sequence 
$U_1,\dots,U_{2|V|-2}\in \mathbf U_G$ such that
\begin{equation}
\ket \psi = U_{2|V|-2} \dots U_1\ket{\varphi}.
\end{equation}
\end{theorem}
\begin{proof}
Suppose $v$ and $w$ are connected vertices, and $\ket \varphi = \alpha_v \ket{v}
+ \alpha_w \ket{w}$. It is simple to show that for arbitrary $\ket{\psi} =
\beta_v \ket{v} + \beta_w \ket{w}$ such that $\braket{\psi}{\psi} =
\braket{\varphi}{\varphi}$, there exists a quantum operation $U$ preserving the
graph such that $\ket \psi = U\ket{\varphi}$. This is equivalent to the fact
that, if $v$ and $w$ are connected, their arbitrary superposition can be
changed to an arbitrary superposition with the same norm.

Let $n=|V|$. The proof goes as follows. First, we show constructively that for
an arbitrary state $\ket{\varphi}$ and vertex $v\in V$ we can find a sequence of
quantum operations preserving the graph structure, changing $\ket{\varphi}$ to
$\ket{v}$. Similarly, one can reverse the method to obtain $\ket{\psi}$ from
$\ket{v}$. In both stages, we need $n-1$ operations. As the result, we obtain a
sequence of length $2n-2$.

Let $T$ be an arbitrary spanning tree of $G$ and $v$ be its arbitrary vertex.
Let $v_1,\dots,v_n$ be an arbitrary order such that if $i>j$, then $d_T(v,v_i) <
d_T(v,v_j)$ (hence $v_n=v$). We define $\suc_T(v_i)= v_j$ such that $j>i$
and $v_i$ and $v_j$ are connected in $T$ for $i\in{1,\dots,n}$, and
$\suc(v_n)=v_n$. Since $T$ is a tree, function is well-defined. Moreover, $\mathbf{U}_T\subseteq \mathbf {U}_G$.
 
Suppose $\ket{\varphi}$ is given. The algorithm goes as follows. At each step
$i\in [n-1]$ we choose quantum operation $U_i \in \mathbf{U}_T$ changing
superposition $\alpha_{v_i}\ket{v_i}+\alpha_{\suc(v_i)}\ket{\suc(v_i)}$ into
$\beta_{\suc(v_i)}\ket{\suc(v_i)}$ such that $|\beta_{\suc(v_i)}|^2=
|\alpha_{v_{i}}|^2+|\alpha_{\suc(v_i)}|^2$ and acting trivially on the other
canonical states. Simple induction shows that, after $n-1$ steps, we obtain state
$\ket{v}$. The reversed method enables us to obtain state $\ket \psi$.
\end{proof}

The above theorem shows that the arbitrary state can be obtained with $O(|V|)$
steps. For some graphs, we need much fewer steps. Even the algorithm from the
proof can be optimised since in some cases the operations can be performed
simultaneously. In the case of the complete graph, an arbitrary state can be
obtained in one step. However, one should note that this bound cannot be
decreased in general. It can be verified by analysing the path graph.

%%%%%%%%%%%%%%%%%%%%%%%%%%%%%%%%%%%%%%%%%%%%%%%%%%%%%%%%%%%%%%%%%%%%%%%%%%%%%%%%
\subsection{Nontriviality of the operations}
%%%%%%%%%%%%%%%%%%%%%%%%%%%%%%%%%%%%%%%%%%%%%%%%%%%%%%%%%%%%%%%%%%%%%%%%%%%%%%%%
While it is straightforward to construct proper stochastic operations preserving
the graph structure, it is not the case for quantum operations. It is because
the construction of stochastic operations can be done independently for each
column. In the case of quantum operations, we need to check whether the columns
are pair-ways orthogonal.

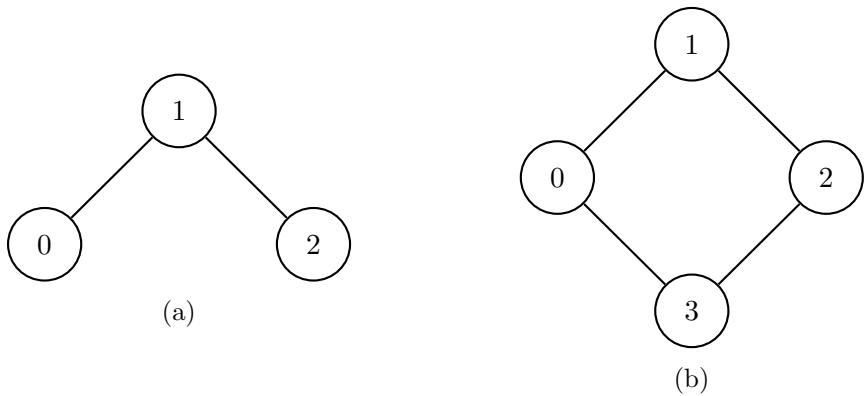
\begin{figure}
  \begin{minipage}{0.45\textwidth}
    \centering
     \begin{tikzpicture}[node distance=2.5cm,thick]
      \tikzset{nodeStyle/.style = {circle,draw,minimum size=2.5em}}
      \node[nodeStyle] (A)  {$0$};
      \node[nodeStyle] (C) [above right of=A] {$1$};
      \node[nodeStyle] (B) [below right of=C] {$2$};
      
      \tikzset{EdgeStyle/.style   = {}}
      \draw[EdgeStyle] (A) to node[left] {} (C);
      \draw[EdgeStyle] (B) to node[right] {} (C);
      \end{tikzpicture}
      \subcaption{}\label{fig:bad-example}
  \end{minipage}
  \begin{minipage}{0.45\textwidth}
    \centering
     \begin{tikzpicture}[node distance=2.5cm,thick]
      \tikzset{nodeStyle/.style = {circle,draw,minimum size=2.5em}}
      \node[nodeStyle] (A)  {$0$};
      \node[nodeStyle] (C) [above right of=A] {$1$};
      \node[nodeStyle] (D) [below right of=A] {$3$};
      \node[nodeStyle] (B) [above right of=D] {$2$};
      
      \tikzset{EdgeStyle/.style   = {}}
      \draw[EdgeStyle] (A) to node[left] {} (C);
      \draw[EdgeStyle] (B) to node[right] {} (C);
      \draw[EdgeStyle] (A) to node[left] {} (D);
      \draw[EdgeStyle] (B) to node[right] {} (D);
      \end{tikzpicture}
      \subcaption{}\label{fig:good-example}
  \end{minipage}
  \caption{(\subref{fig:bad-example}) An example of a graph for which it is
  impossible to find graph-preserving quantum operation changing state state
  $\alpha\ket 0 +\beta \ket 1 + \gamma \ket 2$ to state $\ket 1$. At the same
  time it is simple to find a stochastic operation which performs similar
  operation. (\subref{fig:good-example}) An example of a graph for which it is
  possible to find an operation which changes the state $\alpha\ket 0 +\beta
  \ket 1 + \gamma \ket 2$ to $\ket 1$, for 
$\alpha,\beta,\gamma\neq0$.}
\label{fig:examples}
\end{figure}

Suppose the graph in Fig.~\ref{fig:bad-example} is given. Let us start with
state $\ket \varphi=\alpha\ket 0+\beta \ket 1 + \gamma \ket 2$, where
$\alpha,\beta,\gamma\neq0$. Our goal is to find a single graph-preserving
quantum operation which will change the state into $e^{\ii \psi}\ket 1$. Note
that all quantum operations need to be of the form
\begin{equation}
\begin{bmatrix}
\times & \times  & 0 \\
\times & \times & \times \\
0 & \times & \times 
\end{bmatrix},
\end{equation}
where mark `$\times$' denotes the possibly nonzero value. In order to preserve
the unitarity of the matrix, the first and the last columns need to be
orthogonal. Hence, $\bra 0 A \ket 1$ or $\bra 2A \ket 1$ equals 0. In this case,
the matrix preserves $\alpha\ket0$ or $\gamma\ket 2$ part of the state up to
phase, respectively. The result can be generalized to the star graphs with
number of vertices at least 3. Therefore, it is impossible to obtain the goal.
At the same time, it is easy to find a stochastic operation which changes
arbitrary probability vector $[p_0,p_1,p_2]$ into $[0,1,0]$.

The situation changes diametrically  in the case of the graph in
Fig.~\ref{fig:good-example}. Suppose that we again start in state $\ket
\varphi=r_\alpha e^{\ii k_\alpha}\ket 0+r_\beta e^{\ii k_\beta}\ket 1 + r_\gamma
e^{\ii k_\gamma}\ket 2$, where all the parameters are nonnegative. Then for
\begin{equation}
 V=
\begin{bmatrix}
 -e^{i ({k_\beta}-{k_\gamma}+\psi )} {r_\beta} & e^{i 
 ({k_\alpha}-{k_\gamma}+\psi )} {r_\alpha} & 0 & e^{-i 
 (-{k_\alpha}+{k_\gamma}+\alpha +\pi )} {r_\gamma} \\
 e^{-i ({k_\alpha}-{k_\gamma}+\psi )} {r_\alpha} & e^{-i 
 ({k_\beta}-{k_\gamma}+\psi )} {r_\beta} & e^{-i \psi } {r_\gamma} & 0 \\
 0 & e^{i \psi } {r_\gamma} & -e^{i ({k_\beta}-{k_\gamma}+\psi )} {r_\beta} & 
 e^{-i \alpha } {r_\alpha} \\
 -e^{-i ({k_\alpha}-{k_\gamma}-\alpha )} {r_\gamma} & 0 & e^{i \alpha } 
 {r_\alpha} & e^{-i ({k_\beta}-{k_\gamma}+\psi )} {r_\beta} \\
\end{bmatrix}
\end{equation} 
we have $V\ket{\varphi} =e^{\ii(k_\gamma-\psi)}\ket 1 $. Note that 
\begin{equation}
V\ket 3=
- e^{-i (-k_\alpha+k_\gamma+\alpha  )} r_\gamma \ket 0+
 e^{-i \alpha } r_\alpha \ket 2 +
 e^{-i (k_\beta-k_\gamma+\psi )} r_\beta \ket 3,
\end{equation}
the amplitudes are fixed and we can only change the local phase. 

%%%%%%%%%%%%%%%%%%%%%%%%%%%%%%%%%%%%%%%%%%%%%%%%%%%%%%%%%%%%%%%%%%%%%%%%%%%%%%%%
\section{Quantum controlled Cop and Robber game} \label{sec:the-game}
%%%%%%%%%%%%%%%%%%%%%%%%%%%%%%%%%%%%%%%%%%%%%%%%%%%%%%%%%%%%%%%%%%%%%%%%%%%%%%%%
Taking into account the discussion in Section~\ref{sec:evolution} we can now
define and analyse another quantized version of Cop and Robber game. Using
graph-preserving quantum operations, we generalise the available strategies into
controlled graph-preserving quantum operations. Such game differs significantly
from the previously mentioned ones, at least in the sense of the probability of
winning available for the Cop. We show that a controlled graph preserving
quantum operations generalise the original graph-preserving quantum operations.

%%%%%%%%%%%%%%%%%%%%%%%%%%%%%%%%%%%%%%%%%%%%%%%%%%%%%%%%%%%%%%%%%%%%%%%%%%%%%%%%
\subsection{Model definition}
%%%%%%%%%%%%%%%%%%%%%%%%%%%%%%%%%%%%%%%%%%%%%%%%%%%%%%%%%%%%%%%%%%%%%%%%%%%%%%%%
Similarly to classically controlled quantum Cop and Robber game, each player has
their own quantum system spanned by $\{\ket v : v \in V\}$. In the beginning,
the Cop and the Robber, in that order, choose their initial states. However, in
contrast to the previously defined game, the Robber can entangle arbitrarily
with the Cop at the beginning, by performing the controlled operation. In this
model, the players do not possess the knowledge about the global states and are
allowed to perform controlled graph-preserving quantum operations only. For this
reason, we call this model \emph{quantum controlled Cop and Robber game}.

We define the set of the allowed operations as follows. Let $G$ be an arbitrary
reflexive, connected directed graph. By $\mathbf U_G$ we denote the set of
graph-preserving quantum operations. Moreover, let $U:V\to \mathbf U_G$ be an
arbitrary function. Then the operation
\begin{equation}
\tilde U = \sum_{v\in V} \ketbra{v}{v}\kron U(v)
\end{equation}
is the controlled graph-preserving quantum operation. Note that $\tilde U$ is a
unitary operation. We denote the set of all such operations as
$\mathbf{cU}_G$. One can verify that 
\begin{equation}
\mathbf{cU}_G = \mathbf 
U_{\underbrace{G+G+\cdots+G}_{n\textrm{ times}}},
\end{equation}
where $+$ denotes the disjoint union of the graphs.

In the quantum controlled Cop and Robber game, both players can perform
arbitrary operations from $\mathbf{cU}_G$. The control part is performed on the
opponent's system, while the quantum operations are performed on the player's
system.

\begin{figure}
\begin{center}
\mbox{ \Qcircuit @C=-0.5em @R=-0.5em @! { 
\lstick {\ket{\robber}} & \gate V & \ctrl{1}
& \gate {U_{\robber,1}} &   \ctrl{1} & \gate {U_{\robber,2}} & \qw &
\qcircuitdots & \gate {U_{\robber,t-1}} &  \ctrl{1} & \meter &\cw \\ 
\lstick{\ket{\cop}} & \ctrl{-1} & \gate{U_{\cop,1}} & \ctrl{-1} \qw & \gate
{U_{\cop,2}} & \ctrl{-1} & \qw &\qcircuitdots &  \ctrl{-1} & \gate {U_{\cop,t}}
& \meter &\cw \\ }}
\end{center}
\caption{Quantum controlled Cop and Robber game. 
The Cop chooses his initial state first, then the Robber chooses his state. 
The vertical lines denote quantum control, since both the Robber and the Cop know their and 
their opponent's current state. The measurement is performed in basis $\{\ket v : v 
\in V \}$. $V\in \mathbf U(\Hl_\robber)$ is an arbitrary quantum gate.} 
\label{fig:quantum-controlled-game}
\end{figure}
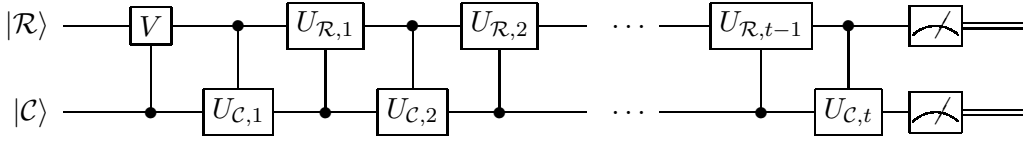

The probability of the Cop to win is the same as in the classically controlled 
quantum Cop and Robber game, \ie{} if after $t$-th round the state of 
$\Hl_\robber \kron \Hl_\cop$ is $\ket{S_t}$, then 
\begin{equation}\label{eqn:pcopwin-ent}
p_{\mathrm{copwin}} = \sum_{v\in V} |(\bra v \kron \bra v)\ket {S_t}|^2.
\end{equation}
However, the formula cannot be simplified due to possible entanglement between
the players' systems. The representation of the game in the form of a quantum
circuit is presented in Fig.~\ref{fig:quantum-controlled-game}. Note that in
this game both players do not know the current global state. For this reason,
both players can prepare the strategies before the game.

The quantum-controlled game expands the classically controlled one in the sense
of the possible operations. Let $U\in \mathbf U_G$. Then
\begin{equation}
\tilde U = \sum_{v\in V}\ketbra{v}{v} \kron U = \left (\sum_{v\in V} 
\ketbra{v}{v} \right ) \kron U = \Id \kron U.
\end{equation}

%%%%%%%%%%%%%%%%%%%%%%%%%%%%%%%%%%%%%%%%%%%%%%%%%%%%%%%%%%%%%%%%%%%%%%%%%%%%%%%%
\subsection{Non-triviality of the game}
%%%%%%%%%%%%%%%%%%%%%%%%%%%%%%%%%%%%%%%%%%%%%%%%%%%%%%%%%%%%%%%%%%%%%%%%%%%%%%%%
The crucial difference between classically controlled and quantum controlled
versions of the quantum Cop and Robber game is observed in the possible set of
strategies. The ability to introduce entanglement between the systems enables
the Cop to win in one step for a large class of graphs.

\begin{theorem}
An arbitrary graph which contains a universal vertex is 1-copwin in quantum 
controlled Cop and Robber game.
\end{theorem}
\begin{proof}
The game goes as follows. The Cop starts in a universal vertex $\ket {v_\cop}$. Since
the Cop starts in the cannonical state, the Robber can only start in an arbitrary,
separated state $\ket{R} = \sum_{v\in V} \alpha_v \ket v$.

Now we define the strategy for the Cop. Let $U_{v\leftrightarrow v'}$ denote a 
transposition matrix, \ie{}
\begin{equation}
U_{v\leftrightarrow v'} = \ketbra{v}{v'} + \ketbra{v'}{v} + \sum_{w\neq 
v,v'}\ketbra{w}{w}.
\end{equation}
If $v,v'$ are connected in $G$, then $U_{v\leftrightarrow v'}\in \mathbf U_G $. 
Obviously, if $v$ is a universal vertex then for an arbitrary vertex $v'$ the 
operation preserves the graph structure. The 
Cop chooses the operation
\begin{equation}
\tilde U_\cop =  \sum_{v\in V} \ketbra{v}{v} \kron U_{v \leftrightarrow v_\cop}.
\end{equation}
The state changes into
\begin{equation}
\tilde{U}_\cop = \left (\sum_{v\in V} \ketbra{v}{v} \kron U_{v \leftrightarrow 
v_\cop} \right ) \ket{R} \kron \ket{v_\cop} = \sum_{v\in V}\alpha_v 
\ket{v}\kron \ket{v}.
\end{equation}
By applying Eq.~\ref{eqn:pcopwin-ent} we obtain the result.
\end{proof}

Note that with the above strategy the Cop wins with probability one in a single
step. The situation differs significantly in comparison to the situation when
both play open probabilistic Cop and Robber game and the classically controlled
quantum Cop and Robber games. In these cases, the Cop wins with probability
$\frac{1}{n}$ only.

Because of the above and the result from \cite{bonato2012almost} we have
the following collorary.

\begin{corollary}
Almost all classical copwin graphs are 1-copwin in the quantum 
controlled Cop and Robber game. 
\end{corollary}

On the other hand it can be seen that the Cop cannot win in the general case.
Let us use a cycle graph $C_4$ as an example. The Cop chooses
$\sum_{i=0}^{3}\alpha_i \ket i$ as the initial state. Then, it is optimal for
the Robber to choose state $\sum_{i=0}^{3}\alpha_i \ket{i,i\oplus 2}$. Simple
analysis shows that for an arbitrary Cop's strategy there always exists a
Robber's strategy such that the state before the Cop's move is of the form
$\sum_{i=0}^{3}\beta_i \ket{i,i\oplus 2}$. Hence, the Cop cannot win the game
with nonzero probability. This example shows that the probability of winning
depends on the graph.

%%%%%%%%%%%%%%%%%%%%%%%%%%%%%%%%%%%%%%%%%%%%%%%%%%%%%%%%%%%%%%%%%%%%%%%%%%%%%%%%
\subsection{Local versus non-local game}
%%%%%%%%%%%%%%%%%%%%%%%%%%%%%%%%%%%%%%%%%%%%%%%%%%%%%%%%%%%%%%%%%%%%%%%%%%%%%%%%

Let us now consider the following unfair game. The Cop is allowed to perform
operations from $\mathbf {cU_{C_4}}$, while the Robber can only perform local
operations from $\mathbf {U_{C_4}}$. Suppose the Cop starts in vertex 
$\frac{1}{2}\sum_{i=0}^3 \ket i$.
Suppose the Robber chooses an arbitrary state $\sum_{i=0}^3 \alpha_i\ket i$. Then 
the Cop chooses the controlled operations described in Section~\ref{sec:evolution}. 
The state is now of the form
\begin{equation}
\sum_{i=0}\sqrt{\frac{3}{4}}\alpha_i \ket{i,i} + \sum_{i,j=0:i\neq j}^{3} 
\gamma_{i,j}\ket{i,j},
\end{equation}
where $\gamma_{i,j}$ comes from the performed operation. Note that
\begin{equation}
p_{\mathrm{copwin}} = \sum_{i=0}^{3} \frac{3}{4} |\alpha_i|^2 = \frac{3}{4}.
\end{equation}
Comparing to the classically-controlled quantum Cop and Robber game, where the Cop can 
achieve the probability $\frac{1}{4}$ at most, the Cop has much better 
possibilities.

%%%%%%%%%%%%%%%%%%%%%%%%%%%%%%%%%%%%%%%%%%%%%%%%%%%%%%%%%%%%%%%%%%%%%%%%%%%%%%%%
\section{Concluding remarks}\label{sec:conclusions}
%%%%%%%%%%%%%%%%%%%%%%%%%%%%%%%%%%%%%%%%%%%%%%%%%%%%%%%%%%%%%%%%%%%%%%%%%%%%%%%%
In this paper, we have introduced the quantum versions of Cop and Robber game.
As a tool, we use quantum operations preserving the structure of the graph. We
show that we can prepare an arbitrary state using such operations for arbitrary
connected, reflexive digraph which contains an undirected spanning tree.
Moreover, we have demonstrated that for arbitrary initial and resulting states
we need a sequence of $2n-2$ operations at most, which is tight in the sense of
complexity. We have also proposed a simple algorithm for obtaining such
sequence.

Using the introduced operations, we quantized Cop and Robber game. We propose
two different quantum models. The classically controlled quantum Cop and Robber
game trivialises in the sense of winning strategies and in this case both the
Cop and the Robber can choose equal superposition as the initial state and
achieve Nash equilibrium. Hence, the probability of winning for the Cop depends
only on the vertex set size and not on any other properties of the graph or the
evolution model. In that sense, the game is similar to open probabilistic Cop
and Robber game. Moreover, we show that both the classically controlled quantum
Cop and Robber game and the open probabilistic Cop and Robber game expand the
original game in the sense of available strategies. In the case of quantum
controlled Cop and Robber game, we allow both players to perform quantum
controlled operations, preserving the graph structure. We argue that the game
differs significantly from both previously defined models in the sense of
winning strategy. By this, we show that the classical control varies
considerably from the quantum control. Moreover, we demonstrate that the quantum
controlled Cop and Robber game expands the original game in the sense of
available strategies.

Unfortunately, we have not found any dependence between the graph-preserving
stochastic operations and the quantum operations.

We have also analysed the case of unfair games. We show that the strategies
available in the open probabilistic and classically controlled quantum Cop and
Robber games are stronger than in the case of deterministic games. We also
demonstrate that the strategies available in the quantum controlled Cop and
Robber game are stronger than in the classically controlled quantum Cop and
Robber game. While in the first one, the result of our analysis is applied to an
arbitrary undirected graph, the latter has been shown by offering some examples.
This analysis demonstrated that the utilization of quantum resources can extend
the space of possible moves. A similar result has been obtained in
\cite{dorbec2017quantum}, where it has been proved that the utilization of
superposition of moves leads to rulesets that may alter the outcomes of games.

The results presented in this paper may be extended in different directions.
First, further analysis of the quantum controlled Cop and Robber game can be
made. Moreover, the various classical generalisations of original Cop and Robber
game can be applied to the quantized version. Further analysis may provide new
information concerning the quantum controlled operations and offer more insight
into the differences between the classical and the quantum versions of
pursuit-evasion games.

%%%%%%%%%%%%%%%%%%%%%%%%%%%%%%%%%%%%%%%%%%%%%%%%%%%%%%%%%%%%%%%%%%%%%%%%%%%%%%%%
\section*{Acknowledgements} 

This work has been partially supported by the Polish National Science Centre
under the project number 2011/03/D/ST6/00413. Authors would like to thank
M.~Ostaszewski for interesting remarks concerning quantum games on directed
graphs.

%%%%%%%%%%%%%%%%%%%%%%%%%%%%%%%%%%%%%%%%%%%%%%%%%%%%%%%%%%%%%%%%%%%%%%%%%%%%%%%
%\bibliographystyle{ieeetr}
%\bibliography{quantum_cops_and_robbers_game}

\begin{thebibliography}{10}
    
    \bibitem{miszczak2012high}
    J.~A. Miszczak, ``High-level structures for quantum computing,'' {\em 
    Synthesis
        Lectures on Quantum Computing}, vol.~4, no.~1, pp.~1--129, 2012.
    \newblock \doiLink{10.2200/S00422ED1V01Y201205QMC006}.
    
    \bibitem{quilliot1978jeux}
    A.~Quilliot, {\em Jeux et pointes fixes sur les graphes}.
    \newblock PhD thesis, Ph. D. Dissertation, Universit{\'e} de Paris VI, 1978.
    
    \bibitem{nowakowski_vertex--vertex_1983}
    R.~Nowakowski and P.~Winkler, ``Vertex-to-vertex pursuit in a graph,'' {\em
        Discrete Mathematics}, vol.~43, no.~2-3, pp.~235--239, 1983.
    \newblock \doiLink{10.1016/0012-365X(83)90160-7}.
    
    \bibitem{bernhard1987rabbit}
    P.~Bernhard, A.-L. Colomb, and G.~Papavassilopoulos, ``Rabbit and hunter 
    game:
    Two discrete stochastic formulations,'' {\em Computers \& Mathematics with
        Applications}, vol.~13, no.~1-3, pp.~205--225, 1987.
    \newblock \doiLink{10.1016/0898-1221(87)90105-2}.
    
    \bibitem{konstantinidis2016simultaneously}
    G.~Konstantinidis and A.~Kehagias, ``Simultaneously moving cops and 
    robbers,''
    {\em Theoretical Computer Science}, vol.~645, pp.~48--59, 2016.
    \newblock \doiLink{10.1016/j.tcs.2016.06.039}.
    
    \bibitem{bonato2016probabilistic}
    A.~Bonato, D.~Mitsche, X.~P{\'e}rez-Gim{\'e}nez, and P.~Pra{\l}at, ``A
    probabilistic version of the game of zombies and survivors on graphs,'' {\em
        Theoretical Computer Science}, vol.~655, pp.~2--14, 2016.
    \newblock \doiLink{10.1016/j.tcs.2015.12.012}.
    
    \bibitem{fitzpatrick2016deterministic}
    S.~Fitzpatrick, J.~Howell, M.~Messinger, and D.~Pike, ``A deterministic 
    version
    of the game of zombies and survivors on graphs,'' {\em Discrete Applied
        Mathematics}, vol.~213, pp.~1--12, 2016.
    \newblock \doiLink{10.1016/j.dam.2016.06.019}.
    
    \bibitem{fitzgerald1979princess}
    C.~H. Fitzgerald, ``The princess and monster differential game,'' {\em SIAM
        Journal on Control and Optimization}, vol.~17, no.~6, pp.~700--712, 
        1979.
    \newblock \doiLink{10.1137/0317049}.
    
    \bibitem{alpern2008princess}
    S.~Alpern, R.~Fokkink, R.~Lindelauf, and G.-J. Olsder, ``The “princess and
    monster” game on an interval,'' {\em SIAM Journal on Control and
        Optimization}, vol.~47, no.~3, pp.~1178--1190, 2008.
    \newblock \doiLink{10.1137/060672054}.
    
    \bibitem{boyer2013cops}
    M.~Boyer, S.~El~Harti, A.~El~Ouarari, R.~Ganian, T.~Gaven{\v{c}}iak, 
    G.~Hahn,
    C.~Moldenauer, I.~Rutter, B.~Th{\'e}riault, and M.~Vatshelle,
    ``Cops-and-robbers: remarks and problems,'' {\em Journal of Combinatorial
        Mathematics and Combinatorial Computing}, vol.~85, 2013.
    
    \bibitem{frankl1987cops}
    P.~Frankl, ``Cops and robbers in graphs with large girth and {Cayley} 
    graphs,''
    {\em Discrete Applied Mathematics}, vol.~17, no.~3, pp.~301--305, 1987.
    \newblock \doiLink{10.1016/0166-218X(87)90033-3}.
    
    \bibitem{chung2011search}
    T.~H. Chung, G.~A. Hollinger, and V.~Isler, ``Search and pursuit-evasion in
    mobile robotics,'' {\em Autonomous robots}, vol.~31, no.~4, p.~299, 2011.
    \newblock \doiLink{10.1007/s10514-011-9241-4}.
    
    \bibitem{bonato2013graph}
    A.~Bonato and B.~Yang, ``Graph searching and related problems,'' in {\em
        Handbook of Combinatorial Optimization}, pp.~1511--1558, Springer, 2013.
    \newblock \doiLink{10.1007/978-1-4419-7997-1\_76}.
    
    \bibitem{bonato2007pursuit}
    A.~Bonato, P.~Pra{\l}at, and C.~Wang, ``Pursuit-evasion in models of complex
    networks,'' {\em Internet Mathematics}, vol.~4, no.~4, pp.~419--436, 2007.
    \newblock \doiLink{10.1080/15427951.2007.10129149}.
    
    \bibitem{bonato2011game}
    A.~Bonato and R.~J. Nowakowski, {\em The game of cops and robbers on 
    graphs},
    vol.~61.
    \newblock American Mathematical Society Providence, 2011.
    \newblock ISBN: ISBN 978-0-8218-5347, \doiLink{10.1090/stml/061}.
    
    \bibitem{piotrowski2003invitation}
    E.~W. Piotrowski and J.~S{\l}adkowski, ``An invitation to quantum game
    theory,'' {\em International Journal of Theoretical Physics}, vol.~42, 
    no.~5,
    pp.~1089--1099, 2003.
    \newblock \doiLink{10.1023/A:1025443111388}.
    
    \bibitem{eisert1999quantum}
    J.~Eisert, M.~Wilkens, and M.~Lewenstein, ``Quantum games and quantum
    strategies,'' {\em Physical Review Letters}, vol.~83, no.~15, p.~3077, 1999.
    \newblock \doiLink{10.1103/PhysRevLett.83.3077}.
    
    \bibitem{pawela2013cooperative}
    {\L}.~Pawela and J.~S{\l}adkowski, ``Cooperative quantum parrondo’s games,''
    {\em Physica D: Nonlinear Phenomena}, vol.~256, pp.~51--57, 2013.
    \newblock \doiLink{10.1016/j.physd.2013.04.010}.
    
    \bibitem{hao2013novel}
    W.~Hao, D.~Chao-Wei, and F.~Bao-Fu, ``A novel multi pursuers-one evader game
    based on quantum game theory,'' {\em Information Technology Journal},
    vol.~12, no.~12, p.~2358, 2013.
    \newblock \doiLink{10.3923/itj.2013.2358.2365}.
    
    \bibitem{miszczak2014quantum}
    J.~A. Miszczak and P.~Sadowski, ``Quantum network exploration with a faulty
    sense of direction,'' {\em Quantum Information \& Computation}, vol.~14,
    no.~13-14, pp.~1238--1250, 2014.
    
    \bibitem{sadowski2016lively}
    P.~Sadowski, J.~A. Miszczak, and M.~Ostaszewski, ``Lively quantum walks on
    cycles,'' {\em Journal of Physics A: Mathematical and Theoretical}, vol.~49,
    no.~37, p.~375302, 2016.
    \newblock \doiLink{10.1088/1751-8113/49/37/375302}.
    
    \bibitem{dorbec2017quantum}
    P.~Dorbec and M.~Mhalla, ``Quantum combinatorial games,'' in {\em 14th
        International Conference on Quantum Physics and Logic (QPL 2017)}, 2017.
    \newblock arXiv:1701.02193.
    
    \bibitem{condon1990algorithms}
    A.~Condon, ``On algorithms for simple stochastic games,'' in {\em Advances 
    in
        computational complexity theory}, pp.~51--72, 1990.
    
    \bibitem{montanaro2007quantum}
    A.~Montanaro, ``Quantum walks on directed graphs,'' {\em Quantum 
    Information \&
        Computation}, vol.~7, no.~1, pp.~93--102, 2007.
    
    \bibitem{bonato2012almost}
    A.~Bonato, G.~Kemkes, and P.~Pra{\l}at, ``Almost all cop-win graphs contain 
    a
    universal vertex,'' {\em Discrete Mathematics}, vol.~312, no.~10,
    pp.~1652--1657, 2012.
    \newblock \doiLink{10.1016/j.disc.2012.02.018}.
    
\end{thebibliography}
%%%%%%%%%%%%%%%%%%%%%%%%%%%%%%%%%%%%%%%%%%%%%%%%%%%%%%%%%%%%%%%%%%%%%%%%%%%%%%%%

\end{document}